# A Generic Storage API

**Graham N.C. Kirby, Evangelos Zirintsis, Alan Dearle & Ron Morrison**
School of Computer Science
University of St Andrews
St Andrews, Fife KY16 9SS, Scotland
{graham, vangelis, al, ron}@dcs.st-and.ac.uk

## Abstract

We present a generic API suitable for provision of highly generic storage facilities that can be tailored to produce various individually customised storage infrastructures. The paper identifies a candidate set of minimal storage system building blocks, which are sufficiently simple to avoid encapsulating policy where it cannot be customised by applications, and composable to build highly flexible storage architectures. Four main generic components are defined: the *store*, the *namer*, the *caster* and the *interpreter*. It is hypothesised that these are sufficiently general that they could act as building blocks for any information storage and retrieval system. The essential characteristics of each are defined by an interface, which may be implemented by multiple implementing classes.

## 1 Introduction

It is increasingly recognised that the traditional approach to software system building, in which fixed abstract components or layers are encapsulated to encourage software reuse, is overly restricting for many applications. The problem is that such fixed software boundaries require the early fixing of policy decisions, which are thus necessarily made to suit the predicted requirements of "typical applications". The policies are then hidden from the application, even though the application may have vital information about which policies are best suited to its needs.

There are various approaches to opening up such restrictions in a controlled manner, so that where appropriate an application may exert control on the policies operated by the underlying software platform. Here we assume that such a mechanism is available, and focus on one particular area of functionality: that of storage.

Previously we have been involved in building a number of object stores—including Napier88 (Morrison, Connor, Kirby, Munro, Atkinson, Cutts, Brown and Dearle, 1999), CASPER (Vaughan, Schunke, Koch, Dearle, Marlin and Barter, 1992), Flask (Munro, Connor, Morrison, Scheuerl and Stemple, 1994) and Lumberjack (Hulse, Dearle and Howells, 1999). Here we are interested in identifying basic storage abstractions that are sufficiently simple and generic to avoid encapsulating particular policies to any significant degree. These abstractions could then be used as building blocks in the construction of various individually customised storage infrastructures. This paper proposes an API embodying one possible set of primitive storage abstractions.

### 1.1 Context

It is straightforward to identify a number of desirable properties of storage systems:

- unbounded capacity
- zero latency
- zero cost
- total reliability
- location independence
- no unauthorised access
- provision of historical views

This set of properties is, of course, a Utopian dream that is never realisable and can only be approximated. Thus storage implementers are faced with a series of technological challenges to meet the aspirations of users. For example, unbounded capacity may be approximated by utilising free space on the network, and zero latency may be approximated by parallel access and caching. This assumes, of course, that data and systems can be organised appropriately to make use of available resources without imposing undue complexity on the user.

In this work we have taken a more limited view, considering the following aspirations:

- Actors, whether users or individual processes, should be able to bind to, update and manipulate data and programs transparently with respect to their respective locations. Thus a given program should work anywhere (with the appropriate infrastructure installed), regardless of its physical location or that of the data accessed. The program should not need to be aware of its own physical location or that of the data accessed.
- Similarly, programs should be expressed independently of the storage and network technology involved in their execution.
- Storage facilities should be structure-neutral: they should not impose their own structure on the information stored. Actors should be able to impose multiple interpretations over information, simultaneously and safely.
- Information should not be discarded; arbitrary historical views should be supported, so that actors may reconstruct information extant at any previous time.
- Protection and security should not be enforced by restricting access to particular information

based on user authentication. Rather, raw stored information should be open to all; where restrictions on its use are required this should be achieved using cryptographic techniques.

Although it was clearly not feasible to meet these aspirations completely, they served as a useful focus in guiding exploration of the various possibilities. The methodology followed was to design a small set of orthogonal components, specified by well-defined interfaces, which could form the building blocks for various storage architectures.

The key advances of the research were:

- the identification of a candidate set of minimal storage system building blocks, which are sufficiently simple to avoid encapsulating policy where it cannot be customised by applications, and composable to build highly flexible storage architectures
- insight into the nature of append-only storage components, and the issues arising from their application to common storage use-cases

## 2 Related Work

The *compliant systems architecture* approach is to separate policy from mechanism wherever possible (Morrison, Balasubramaniam, Greenwood, Kirby, Mayes, Munro and Warboys, 2000). Each component's functionality is delivered by a set of mechanisms, and the policy for using these mechanisms can be supplied by components at conceptually higher levels. In the context of the work described here, we wish to provide storage facilities that are *compliant* to the needs of particular applications. The storage mechanisms should be made available to applications without forcing on them any particular set of policies for their use.

The *open implementation* approach also aims to expose as much policy decision as the applications require, but no more. Techniques include the provision of reflective middleware, allowing inspection and adaptation of the middleware's components (Duran-Limon and Blair, 2002), and meta-object protocols (Kiczales, Lamping, Lopes, Maeda, Mendhekar and Murphy, 1997). Either of these could be used to allow applications to select from a range of storage facilities composed from the primitives introduced here, or to define their own.

The basic storage abstraction proposed here offers append-only storage without update or deletion. This is motivated by work on the log-structured object store known as Lumberjack (Hulse, Dearle and Howells, 1999), which is based on the store technology employed within the persistent operating system Grasshopper (Rosenberg, Dearle, Hulse, Lindström and Norris, 1996). A unique contribution of the Lumberjack store is its non-destructive update of both data and address maps, which allows historical views of the store to be provided to users. Furthermore, the store allows multiple logical logs to be superimposed on a single physical log to facilitate concurrent update.

A number of projects address the provision of storage facilities using peer-to-peer overlay networks. These include OceanStore (Kubiatowicz, Bindel, Chen, Czerwinski, Eaton, Geels, Gummadi, Rhea, Weatherspoon, Weimer, Wells and Zhao, 2000), Mnemosyne (Hand and Roscoe, 2002), PAST (Rowstron and Druschel, 2001b), Pastry (Rowstron and Druschel, 2001a), FreeHaven (Dingledine, Freedman and Molnar, 2001) and Freenet (Clarke, Sandberg, Wiley and Hong, 2000).

Recently efforts have been made to identify a common API to facilitate comparison of such overlay networks (Dabek, Zhao, Druschel, Kubiatowicz and Stoica, 2003). The motivation is similar to that of the work described here, although their design differs in supporting deletion as a primitive operation.

## 3 Proposed Storage API

### 3.1 Generic Components

Four main generic components are proposed: the *store*, the *namer*, the *caster* and the *interpreter*. It is hypothesised that these are sufficiently general that they could act as building blocks for any information storage and retrieval system. The essential characteristics of each are defined by an interface, which may be implemented by multiple implementing classes.

#### 3.1.1 Stores

A *store* component allows arbitrary *bit-strings* to be inserted and later retrieved. No assumptions are made about the format or length of the bit-strings. So that a bit-string may be retrieved, a *key* is returned by the store on its insertion. A key is itself an arbitrary bit-string. All stores implement the following interface:

```
interface Store {
   put:    BitString -> Key
   get:    Key -> BitString  // may fail
   getStoreID: -> BitString
}
```

**Fig. 1: Store Interface**

The *put* operation inserts a given bit-string into the store, and returns a key. If that key is later presented via the *get* operation, the original bit-string is returned. The get operation fails if presented with an unknown key. There are no update or deletion operations, thus a store may be viewed as a monotonically increasing set of key—bit-string bindings. This property was deliberately chosen to make stores suitable as the fundamental building blocks for a storage system with a full historical archive. Where the effects of update and deletion are required by an application, these may be obtained using *namers* as described later.

The policy for key generation is under control of individual stores. Possible policies include: creating keys containing random bit sequences, with sufficient length ensuring low enough probability of accidental clashes; creating keys containing numbers within an increasing sequence; and creating keys by hashing on the content of the bit-strings being stored. Again, with suitable lengths the probability of accidental clashes can be reduced to negligible levels. Using a hashing scheme would open the possibility of information being shared between stores, allowing a data item to be retrieved from a different store from that in which it was originally inserted, since the scheme would ensure that all stores involved mapped the same key to the same bit-string. Thus a store need not necessarily be a *container* for information, so long as it allows insertion and retrieval through the standard interface.

Fig. 2 illustrates the use of the *put* and *get* operations to add a bit-string and later retrieve it.

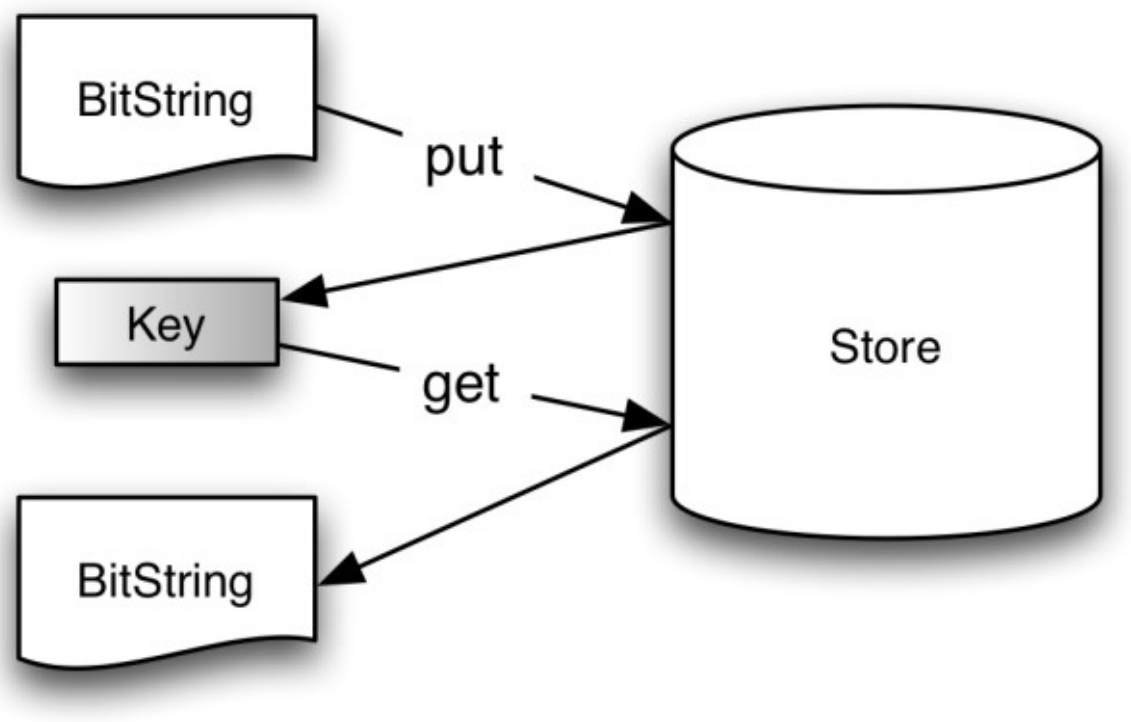

**Fig. 2: Main Store Operations**

The *getStoreID* operation returns a bit-string that is, with high probability, unique to that store instance. This provides a mechanism for encoding references to other stores within a given store. One application for this is the proxy store implementation described later.

A desire for simplicity drove the decision to have a store generate the key for a given bit-string, rather than let the key be supplied by the caller. It could be argued, however, that this departs from the other main motivating principle, that of avoiding the encapsulation of policy, in that the caller cannot control the key generation policy. This design also results in additional complexity when it comes to storing cyclic data structures. For these reasons it might be preferable to add a second variant of the *put* operation with an explicit key:

```
put:    BitString, Key
```

### 3.1.2 Casters

A *caster* component translates information for storage into a bit-string representation suitable for insertion into a store, and vice-versa. A particular caster may be generic, thus applicable to a range of entities, or specific to a particular type of entity. For example, a generic caster has been defined for programming language objects, and specific casters for MS Word documents and XML documents. All casters implement the following interface, where *t* is the type over which the caster operates:

```
interface Caster[t] {
    reify: t -> BitString
    reflect: BitString -> t  // may fail
}
```

**Fig. 3: Caster Interface**

Here it is assumed that *t* may encompass a range of subtypes. The *reify* operation translates a given entity into a bit-string representation. The *reflect* operation performs the inverse, taking a bit-string representation and returning the represented entity. This will fail if presented either with an intrinsically invalid representation, or with a representation for an entity that is not of type *t*. If appropriate, a caster may use cryptographic techniques to verify that a presented bit-string has not been tampered with, and that it did originate from a reified entity of the correct type.

### 3.1.3 Interpreters

An *interpreter* maps one bit-string to another, and may encompass arbitrary computation. Typical uses are for encryption and compression. All interpreters implement the following interface:

```
interface Interpreter {
    interpret: BitString -> BitString
}
```

**Fig. 4: Interpreter Interface**

### 3.1.4 Namers

The components described above are sufficient to allow information of any kind to be stored and retrieved. For practical use, however, two further abilities are required:

- to support update and deletion operations, even though the underlying storage components never discard information;
- to be able to access stored information through symbolic names as well as arbitrary system-specified keys.

These are provided by a *namer* component, which implements a modifiable many-to-many mapping between symbolic names and keys. A name may be bound to multiple keys, allowing a set to be retrieved in a single operation; a key may be bound to multiple names, giving aliasing. Mappings may be updated so that a given name may refer to various keys over time. All namers implement the following interface:

```
interface Namer {
    bind:   Name, Key
    unbind: Name, Key
    lookup: Name -> set[Key]
}
```

**Fig. 5: Namer Interface**

The *bind* and *unbind* operations establish and remove a binding between the given name and key respectively. The *lookup* operation returns all the keys currently bound to the given name; this may be an empty set.

Fig. 6 illustrates the use of the *bind* and *lookup* operations to add a name to key binding, and later to retrieve the set of keys currently bound to that name.

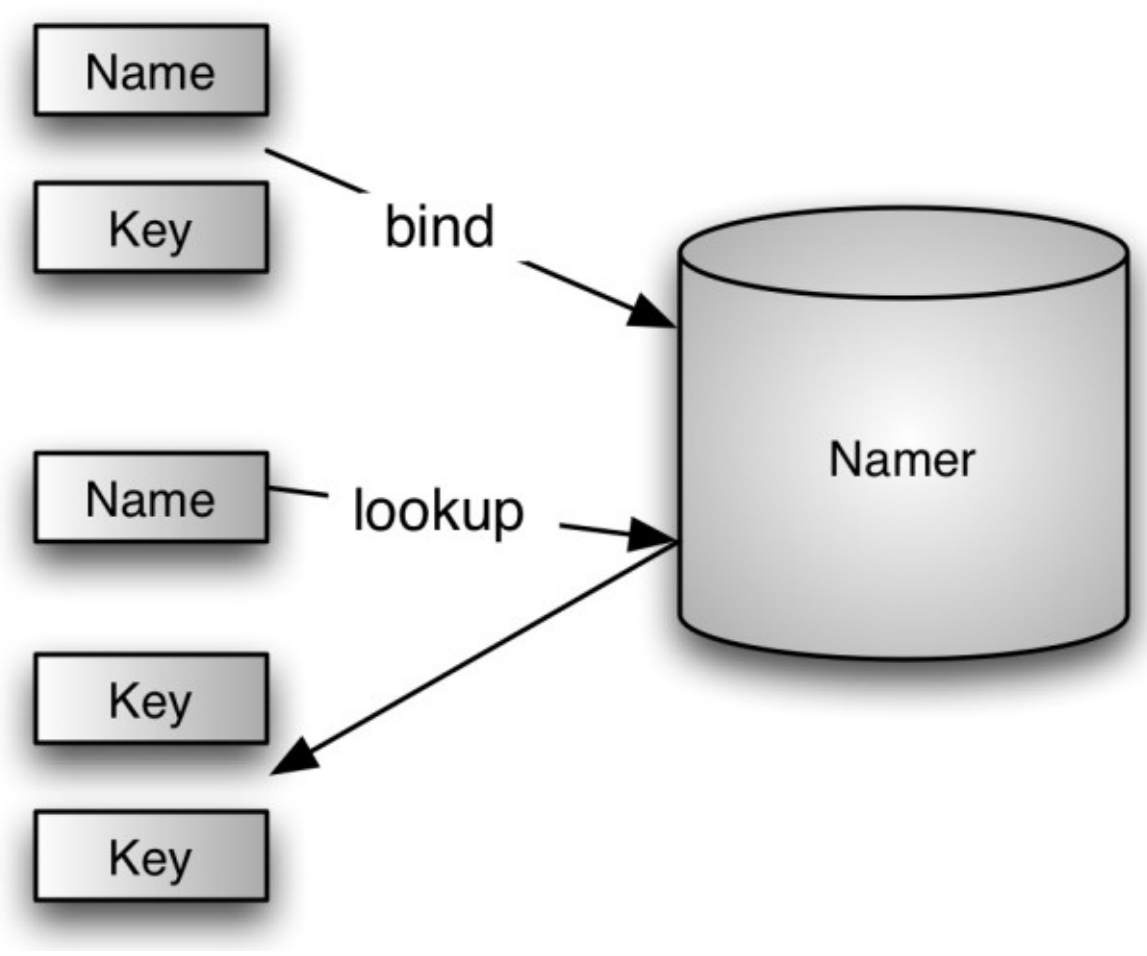

**Fig. 6: Main Namer Operations**

An update operation may thus be provided with respect to symbolic names: a name *n* may be initially bound to a key *k1* using a particular namer; when presented to an appropriate store *k1* allows a data item *d1* to be retrieved. The *n-k1* binding may then be removed from the namer and a new binding between *n* and a key *k2* established. The key *k2* allows the retrieval of a different data item *d2* from the store. Thus the overall effect is an update of the data item corresponding to the name *n*, even though the initial data item *d1* is never discarded from the store.

## 3.2 Implementation

Various implementations of the generic components were developed, and a number of implementation dimensions identified.

### 3.2.1 Stores

Two styles of *store* component were implemented. A *local store* is confined to a single address space on one host machine, and holds its data on that node. A *proxy store* is able to communicate with other stores, both local and remote, and to forward insertion and retrieval A local store may be transient, with its data held solely in memory, or it may have the ability to make its data persistent. One persistent variant appends all inserted bit-strings to a single file, while another creates a new file for each new bit-string, with the file name corresponding to the key allocated to that data item.

A proxy store maintains a set of references to other stores that may be contacted. To enable this to be manipulated the store provides the operations *addTarget* and *removeTarget*. The former adds a new store to the set; this may be specified as a direct reference to another store in the same address space, or as a remote reference to a store on another node, in the form of a URL or a unique identifier. For both local stores and proxy stores it is possible to specify whether a store allows itself to be contacted by other proxy stores. Fig 7 illustrates a proxy store that is connected to two other remote stores. The proxy store functionality could also be used to construct richer topologies such as peer-to-peer networks.

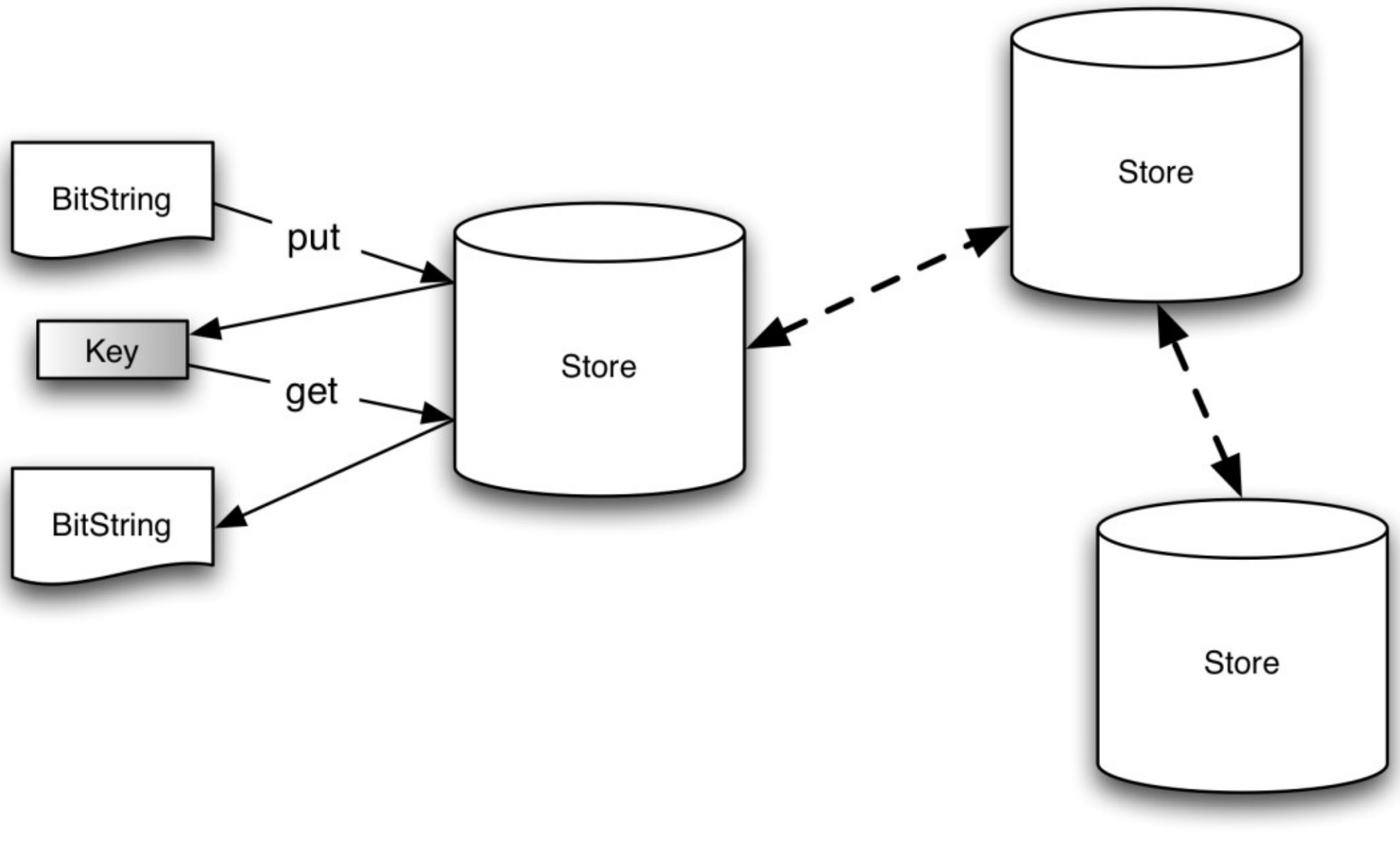

**Fig. 7: Linked Proxy Stores**

To address the bootstrap problem—how does an actor obtain access to an existing store at the start of execution—a static operation *getRootStore* returns a personal store specific to that actor. This root store is persistent in that it organises the storage of its data on non-transient storage. It may itself be either a local store or a proxy store.

### 3.2.2 Casters

Various *caster* components were implemented, including one specifically for XML documents, a generic caster for Java objects, and a caster for store and namer components. The store caster enables a store instance to be reified as a bit-string and stored within another store, thus the *reify-store* sequence of operations may be applied recursively to stores themselves. The store caster operates by reifying a store's contents to a single bit-string in XML form.

#### 3.2.3 Interpreters

Simple *interpreter* components were implemented, providing encryption and compression.

#### 3.2.4 Namers

As with stores, a *namer* component may be transient, in which case it may be saved by reifying it and storing the resulting bit-string in a store, or persistent. In the latter case the namer organises the recording of its bindings on non-transient file storage.

It is also necessary to provide access to a root namer for each actor, obtained via a static operation in the same way as for root stores.

### 3.3 Examples of Use

In order to evaluate the applicability of the generic components in constructing various storage architectures, a number of storage *use cases* were identified, representative of common storage paradigms. The project report (Zirintsis, Kirby, Dearle and Morrison, 2003) contains a full description; a single example is given here for illustrative purposes, showing how the components may be used to store and retrieve a programming language object. The code fragment in Fig. 8 shows an object being reified and inserted into a store.

```
// Create an instance of class Person
Person graham =
   new Person("Graham", 37);

// Retrieve the root store
Store rootStore = XBase.getRootStore();

// Create a Caster for persons
PersonCaster personCaster =
   new PersonCaster();

// Flatten the person into a bit-string
BitString personRep =
   personCaster.reify(graham);

// Put the representation in root store
Key grahamKey =
   rootStore.put(personRep);
```

**Fig. 8: Inserting Data into a Store**

The code fragment in Fig. 9 shows the object being retrieved from a different context:

```
// Retrieve the root store
Store rootStore = XBase.getRootStore();

// Retrieve the representation of the
person using the key
BitString grahamRep =
   rootStore.get(grahamKey);

// Create a Caster for persons
PersonCaster personCaster =
   new PersonCaster();

// Recreate the object
Person reflectedGraham =
   personCaster.reflect(grahamRep);
```

**Fig. 9: Retrieving Data from a Store**

A caster specific to XML documents was developed. In some senses this is not necessary, since a document's textual representation can already be viewed as a bit-string. However, it would not be particularly useful to provide a storage system that simply stored each XML document as a single bit-string in its own file, and thus a scheme to break down documents into multiple bit-strings was implemented.

The XML caster allows the user to determine the granularity of the fragments into which a document is split. At one extreme of the spectrum, a single fragment can contain the entire document, while at the other a separate fragment can be generated for each XML tag. Different points on this spectrum exhibit different tradeoffs with respect to storage space required, overhead in scanning the document, accessing particular regions within it, etc.

Various notations for expressing this granularity were experimented with. The most flexible allows the user to specify a simplified XML schema to which the document conforms; the amount of detail given in the schema determines the fragmentation granularity, and allows control over which sub-parts of the document are reified together within the same bit-string. A simple graphical tool was provided to ease the task of creating this schema, making it relatively straightforward to specify a simple fragmentation pattern: the user simply collapses a sub-tree within the schema in order to specify that corresponding sub-parts in the document should be reified together. This is illustrated in Fig. 10.

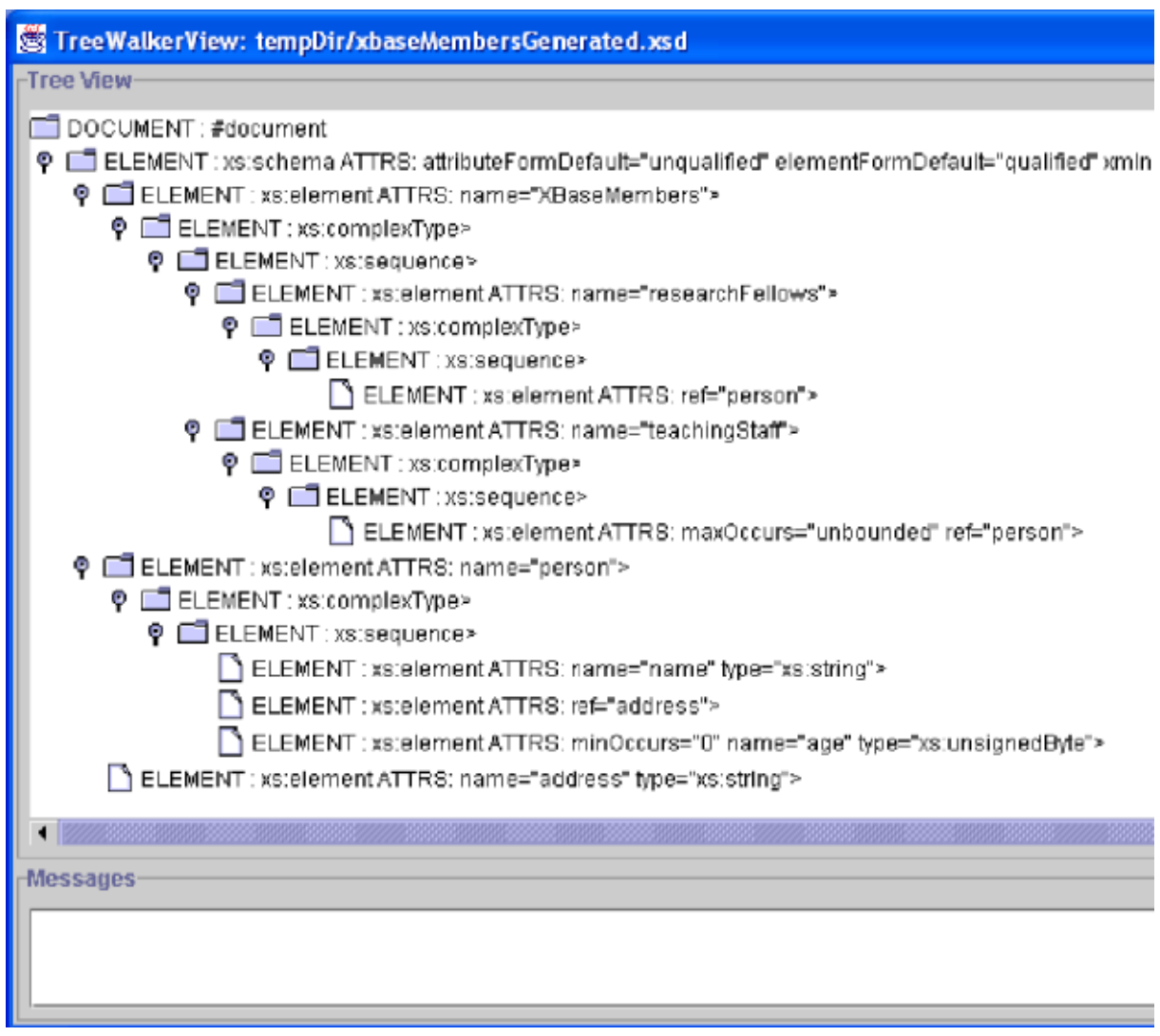

**Fig. 10: Specifying Fragment Granularity**

Other issues include the format of the individual fragments—it was chosen to make these well-formed XML documents in their own right—and the means of representing references between fragments. Inter-fragment references can be represented using keys, where one fragment contains the key corresponding to another fragment, which may be retrieved from an appropriate store. This is simple, but precludes any later update of the document since key-data bindings are fixed. For more flexibility symbolic names may be used, allowing for subsequent modification of the structure. In both cases there is a requirement for some other mechanism to establish which store and/or namer to use. Yet another approach is to make the references fully self-describing, by including some denotation of the appropriate store and/or namer within the reference itself—at the cost of making all references significantly larger.

A further problem with using keys to represent references is that keys are generated within stores rather than externally, so it is not possible to form a reference to a fragment until after that fragment has been inserted into a store and its key obtained. This is not insurmountable for XML, given that the fragment graph is always acyclic, so long as the caster creates the fragments in the correct order. It would be a more significant problem for potentially cyclic structures, as would arise if this approach were to be applied to storage of complete object graphs. In this case, either names would be used to represent references, or, if available, the variant *put* operation described in section 3.1.1 could be used with prior generated keys.

## 4 Further Work

The key advances of this research, as identified earlier, are: the identification of a candidate set of minimal storage system building blocks, and insight into the nature of append-only storage components. Although the tangible results—interface definitions and component implementations—may appear relatively straightforward, these were only arrived at after a prolonged design process that explored a wide range of possibilities.

### 4.1 Autonomic Storage

The research is being continued in a new project “Secure Location-Independent Autonomic Storage Architectures”. This will build on the work described here by further developing the idea of a distributed log-structured (i.e. append-only) storage architecture, in which information may be stored and retrieved transparently with respect to location. The autonomic management aspect (IBM, 2002) will attempt to address the complexity arising from both changing patterns of usage and the various technological opportunities available to the implementer. Infrastructure changes are required to intercept new technologies as they become available. User behaviour changes such as mobility, e.g. working more at home than at work, and software restructuring, e.g. using new or different software, all require complex restructuring of the storage software. User patterns are also influenced by diurnal cycles worldwide; reacting to these patterns efficiently will be essential for high availability. On top of all this, the infrastructure will have to deal with hardware failures.

The project will attempt to relieve the store implementer of the complex tasks of reasoning about computations and resources, allocating, replicating, and moving computations and data to optimise performance, resource usage, and fault-tolerance to meet the desired intrinsic properties. The infrastructure should approximate the Utopian set of ideal characteristics: unbounded capacity; zero latency; zero cost; complete reliability; location independence; a simple interface for users; complete security; and provision of a complete historical archive.

Our approach to engineering a useful approximation involves designing a write-once log-structured storage layer operating above a P2P overlay network. Content-based addressing can be used to achieve location-independent access to data; replication of data “in the right place, at the right time” can be used to achieve reliability and low latency.

Our vision is one of an autonomic storage architecture that presents a simple interface abstracting over all implementation technologies to approximate the user’s desired properties which are extracted automatically from observed usage patterns. This turnkey solution would plug into the user’s chosen operating systems and present a simple view of the store regardless of user location. The aim is to design, implement and evaluate a prototype system of this nature. To achieve this aim we have identified 3 objectives:

- to design a secure location-independent autonomic storage architecture, specified in terms of open interfaces
- to design and implement a corresponding set of plug-compatible components that provide autonomic storage
- to evaluate the architecture and the prototype implementation by deploying it and observing its evolving behaviour under varying loads and usage patterns

Such an architecture is highly dynamic: data flows around the system in response to: changes in users’ location and behaviour; changes in the access patterns of processes; changes in the physical resources allocated to the system; or changes in the topology of the physical infrastructure. It is essential for the underlying policies to evolve in response to such changes, but the complexity is such that it is infeasible for this to be controlled by human users or administrators. The system must therefore be autonomic, managing such changes automatically.

### 4.2 Other Projects

The research is also feeding directly into a number of other ongoing projects. The Cingal project (Dearle, Connor, Carballo and Neely, 2003), a joint project between St Andrews University and Strathclyde University, is developing *thin server* technology to allow code and data to be pushed safely to appropriate locations in a global network. The work described here is influencing the design of storage facilities incorporated into thin servers.

The GLOSS project (Dearle, Morrison, Kirby, Nixon, Connor, Dunlop, Coutaz and Clarke, 2000) seeks to develop a distributed event-based infrastructure to support the deployment of pervasive contextual services on a global scale. A crucial aspect of this is the storage of events and other contextual information on widely distributed nodes, to which the current research will be highly relevant (Dearle, Kirby, Morrison, McCarthy, Mullen, Yang, Connor, Welen and Wilson, 2003) (Kirby, Dearle, Morrison, Dunlop, Connor and Nixon, 2003).

Finally, in ArchWare (Morrison, Kirby and Balasubramaniam, 2001), a project on evolvable software architectures, this research is contributing to thinking on open, software systems that are susceptible to evolution.

## 5 Conclusions

This paper has proposed a simple and generic storage API, which could be exposed directly to applications that have need to exert fine control over storage implementation policies. The initial motivation was flexibility; further experimentation is required to investigate whether the API could be implemented so as to deliver acceptable performance and scalability.

One of the most interesting research questions opened up by this work is the viability of pervasive global storage, accessible from anywhere, from which no information is ever discarded. Intuitively this currently seems unachievable, but continuing research coupled with further advances in storage hardware technology may well allow this ideal to be closely approximated.

## 6 Acknowledgements

This work was supported by EPSRC grant GR/R45154 “Bulk Storage of XML Documents”. Dharini Balasubramaniam and Aled Sage also contributed to the work. Further research in this area is being supported by EPSRC grant GR/S44501 “Secure Location-Independent Autonomic Storage Architectures”.